\documentclass[a4paper,10pt,twoside,twocolumn]{article}
\usepackage{ercoftac_2012}
\usepackage{hyperref}
\hypersetup{
    colorlinks=true,
    linkcolor=blue,
    filecolor=magenta,      
    urlcolor=cyan,
    citecolor=teal,
    }

\usepackage[absolute,showboxes,overlay]{textpos}
\TPGrid[2mm,2mm]{20}{30}
\TPMargin{1.5mm}
\newcommand{\pder}[2]{\frac{\partial #1}{\partial #2}}

\begin{document}

\title{{\vspace{-0.0cm}\sc{On the simulation of multicomponent and multiphase compressible flows}}}
\author{R\'{e}mi Abgrall$^1$, Paola Bacigaluppi$^2$, Barbara Re$^1$\footnote{Corresponding author: \texttt{barbara.re@math.uzh.ch}}\vspace{0.3cm}\\
\small{\it{$^1$Institute of Mathematics, University of Z\"{u}rich}}\\
\small{\it{$^2$ Laboratory of Hydraulics, Hydrology and Glaciology, ETH Z\"{u}rich}}\\}
\date{}
\maketitle
\thispagestyle{fancyplain}
\vspace{0.5cm}

\section*{Abstract}
The following paper presents two simulation strategies for compressible two-phase or multicomponent flows.
One is a full non-equilibrium model in which the pressure and velocity are driven towards the equilibrium at interfaces by numerical relaxation processes, the second is a four-equation model that assumes stiff mechanical and thermal equilibrium between phases or components.
In both approaches, the thermodynamic behaviour of each fluid is modelled independently according to the stiffened-gas equations of state.
The presented methods are used to simulate the de-pressurization of a pipe containing pure CO\textsubscript{2} liquid and vapour under the one-dimensional approximation.

\textblockrulecolour{purple}
\begin{textblock}{15}(5,3) \small
\color{purple}
\textbf{ERCOFTAC Bulletin 124, September 2020. }
\textit{Accepted version}.

\footnotesize
Final publication available at
\href{https://www.ercoftac.org/publications/ercoftac_bulletin/bulletin-125/}{www.ercoftac.org/publications/ercoftac\_bulletin/bulletin-125/}
\end{textblock}

\section{Introduction}
Multiphase and multicomponent flows are ubiquitously encountered in nature as well as in industrial applications. Hence, their numerical simulation has been an active research field for a long time producing a large variety of models as well as numerical methods relying on the most different assumptions.
Among the possible distinctive aspects of such a modelling, is the thermodynamic description. 
In this work, we assume that each (pure) fluid has its own, known, equations of state (EOS), independently from the actual phase---liquid or gaseous---that composes the flow. How the phases or materials interact among each other represents another important feature. Specifically, we focus on the so-called diffuse interface methods (DIMs), which consider in each cell the presence of all phases or materials, be it an arbitrarily small amount for the ``pure'' fluids, and do not explicitly track the interfaces separating different components~\cite{Saurel1999, Saurel2018}.\\

To allow an effective description of the considered numerical methods, it is important to remark the difference between multicomponent and multiphase flows.
In \textit{multicomponent flows}, the different species (typically gases, but also liquid mixtures) are intimately mixed and, whatever the size of a volume of fluid, the different components appear all. An example is given by air, where O\textsubscript{2} and N\textsubscript{2}, as well as the other gases, are simultaneously present, but the mixture composition varies.
In such cases, the Dalton law applies, and the pressure is the sum of the partial pressures.
This kind of flows are mostly considered to simulate gas combustion, hypersonic flows, or real gas networks, where several gases are injected from different origins.
In \textit{multiphase flows}, different phase are separated by multi-material interfaces, so there exists a limit size for which one can say in which pure fluid we are.
In \textit{multiphase flows}, each phase occupies a well-determined volume fraction, i.e., there exists a limit in the spatial scale for which one can say in which pure fluid we are.
Take the example of a fog, where droplets can appear in a gaseous environment: if we are looking below the characteristic size of a droplet of water, then we know if we are in, say, gas or in water. 
Such types of flows are generally assumed while investigating combustion processes, break-up of liquid jets, but also in the oil industry, the nuclear industry, and during the transport of CO\textsubscript{2}, etc..
According to the flow topology, we can have, for instance, stratified, bubbly, or droplet two-phase flows.

From a numerical point of view, these two descriptions have many similar common features, but also many differences.
If we neglect viscous effects, a multicomponent model always leads to a hyperbolic system, and the main difference with the single component case (e.g., with the standard Euler equations), is that the number of linearly degenerate fields increases.
The main difficulty occurs at the contact discontinuities where the pressure may not have the right continuous behaviour that is expected.
This artefact becomes more important if the slip line is aligned with the mesh, and may lead, in some cases, to the blow up of the code.
The case of multiphase flows is way more complicated. 
First, there is a very difficult modelling issue: it is not possible, in general, to have a very fine description of the flow, but we have to introduce averages, a situation somehow similar to turbulence modelling. Hence, we need closure relations, and they are case dependent.
This averaging procedure is a translation, from the physical to the numerical level, of the flow topology;
hence, it depends on the shape of interfaces, distinguishing e.g. between small, large or elongated bubbles, or on the behaviour, e.g. how a bubble breaks. The averaging procedure might impact also the mathematical properties of the model.\\

Here, we intend to describe two possible ways of proceeding that are suitable to simulate both compressible multiphase and multicomponent flows.
The proposed strategies are based on primitive formulations. It means that the set of partial differential equations describing the flows is not written in conservative form, and the solution variables comprise, instead of the total energy, the pressure or the internal energy, which are, in general, more relevant in engineering applications.
One aspect linked with this choice, which requires some attention and it is non-trivial, is the involved numerical method to guarantee conservation among the variables for the four-equation model \cite{abgrall2018high,bacigaluppi2019high}.

In this contribution, we narrow our interests to two-phase flows (or two components, as, in the following, we will not distinguish any more between them). First, in Sec.~\ref{sec2}, we introduce diffuse interface methods for the simulation of compressible two-phase flows.
In Sec.~\ref{s:seveneqs}, we describe a pressure-based finite-volume solver for the seven-equation model, in which the equilibrium between pressure and velocity at multi-material interfaces is enforced through numerical relaxation processes.
In Sec.~\ref{s:foureqs}, we briefly resume the finite-volume-type solver adopted for the four-equation model.
The results of a CO\textsubscript{2} pipe de-pressurization test are presented in Sec.~\ref{s:results}, while the conclusions drawn in Sec.~\ref{s:conc}.

\section{Diffuse interface methods for two-phase flows}
\label{sec2}
A distinguishing feature of compressible two-phase flows is the presence of dynamic interfaces separating different fluids, which represents a main challenge in their numerical simulation.
The size, shape, and the number of interfaces present in the domain depend on the topology of the flows, and the thermodynamic and/or chemical properties of the flow may undergo abrupt variations across them.
DIMs is possibly an easy and efficient way to cope with moving interfaces, as it artificially (i.e., numerically) lets the fluids mix in a thin region surrounding them~\cite{Saurel2018}. In other words, the dynamics of the fluids is coupled in thin interfacial regions, while it tends to their respective pure behaviour away from them.
The modelling of the thermodynamic behaviour of the mixture near interfaces, which may significantly depart from the bulk fluids, is a rather delicate matter and it can be controlled by different modelling assumptions.

The archetype of the DIMs is the full non-equilibrium model by Baer and Nunziato (BN)~\cite{Baer1986}, which comprises seven equations: the evolution equation for the volume fraction of one phase and a set of balance equations for mass, momentum and total energy for each phase. 
Each phase evolves with its own pressure, velocity, and temperature.
The original BN model has been equipped with instantaneous pressure and velocity relaxation, to enforce mechanical equilibrium across interfaces~\cite{Saurel1999}.
Despite the great flexibility and some favourable numerical features such as the hyperbolic character,
the extensive use of BN-like models is hindered by the large number of waves, that need to be taken into consideration~\cite{Tokareva2010}.
This aspect prompts the diffusion of reduced models, which are derived by means of asymptotic expansions, assuming pressure, velocity, and/or thermal equilibrium between phases~\cite{Lund2012}.
Among others, we mention the six-equation single-velocity two-phase flow model~\cite{Saurel2009}, and the five-equation model derived by Kapila et al.~\cite{Kapila2001} in the limit of stiff mechanical relaxation, and the 4-equation homogeneous equilibrium model with mechanical and thermal equilibrium~\cite{downar1996non,LundAursand2012}.
The choice of the most suitable model depends on the features of the flow field, the desired simulation outputs, and the available computational resources.\\

A difficulty that we need to face while developing numerical methods for compressible multiphase flows concerns the non-conservative terms.
In the seven- and six-equation models, these terms result from the averaging procedure and involve the gradient of the volume fraction. In the five-equation model, it involves the velocity divergence.
The impossibility to write the governing equation in conservative form precludes the definition of weak solutions in the standard distribution sense and calls for ad-hoc numerical techniques, which provide an unambiguous discretization of non-conservative terms.
A possible one is provided by the discrete equation method, which applies the principle of the Godunov method in a probabilistic framework~\cite{Abgrall2003}.
Alternative, we can derive a numerical discretization that explicitly enforces the pressure and velocity non-disturbance condition across multi-material interfaces~\cite{Abgrall1996}.
More specifically, it can be noticed that if the two fluids are in mechanical equilibrium, the pressure and the velocity are and should remain constant across the interface.
Enforcing the mechanical equilibrium at multi-material interfaces is mandatory to avoid spurious velocity and pressure oscillations.
Primitive formulations offer a natural way to enforce the interface mechanical equilibrium~\cite{Karni1996}, so we focus on them in this paper.\\

In the next sections, we describe a finite-volume method that solves the pressure-based formulation of the seven-equation non-equilibrium model, with finite pressure and velocity relaxations, where the pressure and velocity non-disturbance constrain is inherently enforced in the finite volume discretization; along this model, we also present a finite-volume-type solver for the non-conservative four-equation model written in terms of internal energy. This system in particular considers pressure and velocity, along with temperature relaxations.

\subsection{A pressure-based non-equilibrium BN-type model}
\label{s:seveneqs}
The non-equilibrium model is based on the symmetric variant of the BN model with pressure and velocity relaxation proposed by Saurel and Abgrall~\cite{Saurel1999}.
With the aim to derive a model well suited to simulate multiphase flows at low Mach numbers, we derived the corresponding pressure-based formulation~\cite{Re2019}, which is made dimensionless according to the special pressure scaling
\begin{equation*}
P_k = \frac{\widetilde{P}_k - \widetilde{P}_\mathrm{r}}{\widetilde{\rho}_\mathrm{r} \widetilde{u}_\mathrm{r}^2}\quad \text{with} \quad
M_\mathrm{r}^2 = \frac{\widetilde{\rho}_\mathrm{r} \widetilde{u}_\mathrm{r}^2}{\widetilde{P}_\mathrm{r}} \,,
\end{equation*}
where $\widetilde{P}_k$ is the dimensional pressure of the phase $k$, while $\widetilde{P}_\mathrm{r}$, $\widetilde{\rho}_\mathrm{r}$, and  $\widetilde{u}_\mathrm{r}$ are, respectively, the reference (dimensional) pressure, density, and velocity.
$M_\mathrm{r}$ is the resulting reference Mach number, which expresses the global level of compressibility of the flow field, as explained later.
This special scaling filters out the acoustics and recovers the correct order of pressure fluctuations in the zero Mach limit~\cite{Wenneker2002}.
The resulting model reads~\cite{Re2020}:
\begin{align}
&\pder{\alpha_k}{t}+ u_\mathrm{I}\pder{\alpha_k}{x}  =\mu \Delta_k P \label{eq:bna}\\
&\pder{\alpha_k \rho_k}{t} +  \pder{(\alpha_k \rho_k u_k)}{x}=0 \label{eq:bnrho}\\
&\pder{\alpha_k m_k}{t} + \pder{(\alpha_k m_k u_k + \alpha_k P_k)}{x} = P_\mathrm{I}\pder{\alpha_k}{x} - \lambda \Delta_k u \\
&\begin{aligned}
M_\mathrm{r}^2 \alpha_k&\biggl[\pder{P_k}{t} +   u_k\pder{P_k}{x}\biggr]  \!\!
  +\bigl[  M_\mathrm{r}^2  \rho_k c_k^2 + \kappa_k \biggr] \alpha_k \pder{u_k}{x}   \\
 - &\biggl[ M_\mathrm{r}^2 \rho_k   c^2_{\mathrm{I},k} +\kappa_k \biggr]  (u_\mathrm{I}\! - u_k)\pder{\alpha_k}{x} \\
 =- M_\mathrm{r}^2&\biggl[  \rho_k c^2_{\mathrm{I},k}  \mu \Delta_k  P \! +   \kappa_k (u_\mathrm{I} \!- u_k) \lambda \Delta_k u \biggl] \! - \kappa_ \mu \Delta_k P 
  \end{aligned}\label{eq:bnP}
\end{align}
where $\alpha$ is the volume fraction, $m$ the momentum, and the subscript $k=\lbrace 1 , 2 \rbrace$ indicates the phase.
The operator $\Delta_k $ takes the difference between the phase $k$ and the opposite one, e.g.,  $\Delta_1 u = u_1 - u_2$, and 
$\lambda$ and $\mu$ are the relaxation parameters for the velocity and the pressure, respectively.
Since the phases occupy all volume, $\alpha_1 + \alpha_2 =1$, so \ref{eq:bna} is solved only for one phase, while \ref{eq:bnrho}--\ref{eq:bnP} are solved for both phases.
The system is closed by two thermodynamic models, one for each phase, which are assumed to be given in the general form $P=P(\rho, e)$, with $e$ the internal energy.
Accordingly, the speed of sound $c$ is defined for each phase as:
\begin{equation} \label{eq:spSound}
c^2 \!= \chi + \kappa \frac{P + e}{\rho},  \quad \text{with}\;\;
\chi = \!\left(\pder{P}{\rho}\right)_{\!\!e}\!\!, \;\; \kappa =\! \left(\pder{P}{e}\right)_{\!\!\rho}\!\!.
\end{equation}
Finally, the subscript $\mathrm{I}$ denotes interfacial variables, which are modelled as
\begin{gather*}
u_\mathrm{I}=\frac{\sum_k \alpha_k m_k}{\sum_k \alpha_k \rho_k} \,, \qquad  P_\mathrm{I}=\sum_k \alpha_k P_k \,, \\
c^2_{\mathrm{I},k} = \chi_k + \kappa_k \frac{P_I + e_k}{\rho_k} \,.
\end{gather*}
The definition of the \textit{interfacial} speed of sound $c_{\mathrm{I},k}$ is not related to a specific EOS, and it does not have a formal thermodynamic meaning, but it simply mimics \ref{eq:spSound}, for the pressure $P_\mathrm{I}$ instead of $P_k$.
The usage of two EOSs circumvents the possible occurrence of negative values of squared speed of sound in the two-phase region~\cite{Saurel1999}.

In \ref{eq:bnP}, we highlight the role of the parameter $M_\mathrm{r}^2$, in front of some terms.
It is representative of the global compressibility of the flow and it allows to recover the correct incompressible solution at low Mach~\cite{Munz2003}.
Indeed, for $M_\mathrm{r}^2 \rightarrow 0$, \ref{eq:bnP} reduces to $\pder{\alpha_k}{t}+ \pder{\alpha_k u_k}{x} =0$,
which can be considered the multiphase counterpart of the incompressible kinetic constraint on the velocity divergence for single phase flows.\\

The solution strategy of the pressure-based seven-equation model given above is based on the Strang splitting, namely, the solution at the end of time step $\Delta t = t^{n+1}- t^n$ is achieved by applying two subsequent operators: the hyperbolic operator, which solves the system without source terms, and the relaxation operator, which solves two ordinary differential equation (ODE) systems associated with the velocity and the pressure relaxation, starting from the solution of the hyperbolic operator.

The numerical discretization of the hyperbolic operator is described in~\cite{Re2019}. We recall here only some core features.
The governing equations are spatially discretized over staggered grids to avoid spurious pressure oscillations at low Mach number, and we use a first-order finite volume scheme based on the Rusanov fluxes. The staggering of the variables facilitates the solution of the system of equations in a segregated approach, so first the densities and the volume fraction and the predicted momentums are computed, by treating explicitly the convective terms and the pressure gradients.
Then, the pressure equations are solved by integrating implicitly the acoustic terms, in order to circumvent the most stringent limitation on the time step imposed by the acoustic CFL constraint.
Finally, the momentums and velocities are updated according to the pressure correction.
As mentioned in the first part of Sec.~\ref{sec2}, the conservative terms involving the gradient of the volume fraction are integrated starting from the consideration that a two-phase flow uniform in pressure and velocity should preserve this uniformity~\cite{Saurel1999}.
By analytically imposing this condition in the discrete version of the equations, we achieve an unambiguous discretization of the non-conservative terms, which, however, depends on the adopted numerical fluxes~\cite{Re2020,Saurel1999}.

For what concerns the relaxation operator, we solve two systems of ODEs, which include only the time derivatives of the variables and the relaxation terms (we refer the reader to~\cite{Saurel1999} for a description of the systems of ODEs, but here $\lambda$ and $\mu$ are finite parameters).
The first system accounts for velocity relaxation only, and it consists of the two momentum equations.
Using a backward Euler time integration scheme and considering the solution of the hyperbolic operator as initial state, we obtain an easy system which is solved analytically.
The second ODEs problem applies the pressure relaxation and it consists of three equations: one for the volume fraction and two for the pressure.
The implicit Euler integration of this initial value problem leads to a highly non-linear problem, which is solved by using standard numerical techniques provided by the library~\textsf{PETSc}~\cite{petsc-web-page}.\\

Finally, we recall that the standard BN-type models, as the one here described, consider immiscible phases, that is, unless the components are not premixed, the only mixing that may occur is due to numerical diffusion.
However, mass transfer may be added to the model by means of additional source terms modelling the Gibbs free energy relaxation~\cite{Zein2010}.

\subsection{An internal energy-based four-equation model}
\label{s:foureqs}
The four-equation model is a simplified version of the Baer-Nunziato two-phase model \cite{Baer1986}, with the assumption of both mechanical and thermal relaxation.
In particular, to retrieve the four equations starting from the Baer-Nunziato model we have assumed pressure and velocity, as well as temperature equality between the two phases (cf. Sec.~\ref{s:thermo} for further details). 
This leads to the following set of equations, where we have one equation for the conservation of mass for each phase $k$ and an equation for the mixture momentum \ref{4eqs_mom} and the mixture internal energy \ref{4eqs_ener}.
 \begin{align}
&\pder{(\alpha_1 \rho_1)}{t}+ \pder{(\alpha_1 \rho_1 \; u)}{x}=0\label{4eqs_mass1}\\
&\frac{\partial (\alpha_2 \rho_2)}{\partial t}+ \pder{(\alpha_2 \rho_2 \;u)}{x}=0 \label{4eqs_mass2}\\
&\frac{\partial ( \rho \;u)}{\partial t}+\pder{\;( \rho\;u^2 +P )}{x}=0 \label{4eqs_mom}\\
&\frac{\partial e}{\partial t}+ u \cdot \pder{ e}{x}  + (e+P)\;\pder{u}{x}=0 \label{4eqs_ener}
\end{align} 
Note that system \ref{4eqs_mass1}-\ref{4eqs_ener} is written for the one-dimensional case and, more specifically, the last equation is featuring the non-conservative formulation, and requires a careful numerical method, in order to provide the conserved quantities.
Further, the thermodynamic assumptions behind this model are different from the original Homogeneous Relaxation Model (HRM) model of Downar \cite{downar1996non}, which does not assume a temperature equilibrium. 

Let us denote the two phases identified for a liquid phase by subscript $1$ and a gaseous one by $2$.
Here $u$ represents the velocity, $\rho$ the mixture density, $P$ the pressure and $c$ the mixture speed of sound. The Wood velocity $\frac{1}{\rho c^2}=\sum_k \frac{\alpha_k}{\rho_k c_k^2}
$ describes the total sound speed, where $c_k^2$ is the squared speed of sound for phase $k$, as defined in~\ref{eq:spSound}.\\
The volume fractions $\alpha_k$ fulfil the requirement
$\sum_k \alpha_k=1.$
We define the total energy as 
$
E=e+\frac{1}{2}\rho u^2,
$
with the  total internal energy $e=\sum_k \alpha_k e_k$, where $e_k$ is the internal energy of phase $k$.\\

The numerical discretization of the four-equation system is described in \cite{bacigaluppi2019high,bacigaluppi2020}. We recall hereafter some of its main features. The governing equations are spatially discretized via a finite volume-based technique, called the Residual Distributions Finite Volume (RD), which has been duly presented in some recent work \cite{abgrall2017design,BacigaluppiMood2018}. This method is used in combination with a second-order expicit (Runge-Kutta) method, which has been inspired by the work of \cite{Ricchiuto2010} and \cite{abgrall2017design}. This temporal discretization is based on a prediction-correction approach, where the prediction step is first-order accurate and given by a flux difference. The second-order in time is then achieved via a correction step, which takes the obtained prediction approximation as the previous sub-timestep solution.

Overall, for what concerns the numerical spatial approximation, diverse methodologies, ranging from Lax-Friedrichs to Godunov approximations, could be adopted.
Nevertheless, to guarantee overall a good accuracy of the solution, providing at same time a robust scheme, has led to the choice of adopting a so-called a posteriori limiting, as presented in \cite{BacigaluppiMood2018}. This strategy allows us to combine a second-order accurate scheme in smooth regions and reverting locally to first-order in case the second-order does not fulfil a series of criteria, such as the absence of numerical oscillations or physically inadmissible solutions, as, e.g., negative densities.
In this work, the second-order approximation is retrieved by a stabilized, blended Rusanov scheme, while the first-order is provided by the Rusanov scheme. The interested reader may refer to \cite{bacigaluppi2019high,bacigaluppi2020}.
While the feature of accuracy can be well preserved, some remarks on the conservation among the variables need to be provided.

In particular, the feature  behind the choice of a non-conservative formulation, as seen in \cite{abgrall2018high}, is due to the many advantages, especially for engineering applications, which are represented by the possibility to work directly with quantities, such as pressures or internal energies instead of having them computed from the  total energy. 
Indeed, for example, across contact discontinuities, the velocity and the pressure do not change, while the density does. Deriving the internal energy or pressures from the total energy, may, therefore, not be completely accurate from a numerical point of view, possibly resulting in oscillations. The interested reader may refer to \cite{bacigaluppi2019high} for a comparisons between conservative and non-conservative formulations.
Further, another relevant advantage is represented by the possibility of dealing easily with non-linear equations of state, as for example the family of Mie-Gr\"uneisen equations of state.
These reasons mainly motivate the interest behind the adoption of \ref{4eqs_ener} and the conservation can be retrieved in the specific RD scenario via a correction term (cf. \cite{abgrall2018high}).
Mass transfer may be added to \ref{4eqs_mass1}-\ref{4eqs_mass2}, on the right-hand side, by means of  the term $\Gamma=\theta(\nu_1-\nu_2)$ with $\nu$ the chemical potentials (Gibbs function) of the phases and $\theta$ the relaxation time of the process at which the thermodynamic equilibrium is reached (which enables the fulfilment of the entropy inequality).
The solution would then be obtained by first computing the hyperbolic part and then proceeding by adding the mass transfer as shown in \cite{Pelanti2014,bacigaluppi2019high,bacigaluppi2020}.

\subsection{Thermodynamic modelling}
\label{s:thermo}
In this work, we consider a stiffened gas EOS for each phase. Considering dimensional variables, the pressure and the temperature are given by 
\begin{gather*}
P(\rho, e) = (\gamma - 1) e - \gamma P_\infty - (\gamma -1 )\rho q \,,\\
T(\rho, e) = \frac{P(\rho, e) + P_\infty}{(\gamma - 1 )c_v \rho}\,,
\end{gather*}
where the parameters $\gamma$, $c_v$, and $q$ are, respectively, the ratio of specific heats, the isochoric specific heat capacity, and the zero point energy.
Finally, the parameter $P_\infty$ models the attractive effects in condensed materials.
All these parameters are fluid-specific and are computed by fitting experimental data, e.g. the saturation curve~\cite{LeMetayer2004}.

Despite its simple analytical formulation, the stiffened EOS permits to model the essential physics at molecular level, that is attractive and repulsive effects. Hence, it is often used for shock dynamics in condensed materials and to simulate phase transfer.\\

For completeness, we briefly mention how the temperature, pressure and velocity equality among phases is reached for the four-equations, having possibly different states, as e.g. densities.
For the temperature we will have   \begin{equation*}
T=\sum_k \frac{P+P_{\infty,k} }{ \alpha_k\rho_k{C_v}_k\; (\gamma_k -1)  },
\end{equation*}
whereas for
the internal energy $$
e=\sum_k \alpha_k \rho_k \Big{(}\,{C_v}_k\; T \frac{P +\gamma_k\; P_{\infty,k}}{P+P_{\infty,k}}+\;q_k \Big{)}.
$$ and, finally, for the
pressure \begin{equation*}
\begin{split}
P=&\frac{1}{2}  \sum_k \left(A_k -P_{\infty,k}\right)\\&+\sqrt{\frac{1}{4}(A_2-A_1-(P_{\infty,2}-P_{\infty,1}))^2+A_1\;A_2}\,,\end{split}
\end{equation*}
with
$$
A_k=\frac{\rho_k\alpha_k(\gamma_k-1) C_{v_{k}}}{\sum_k\rho_k\alpha_kC_{v_k}} \Big{(} e-\sum_k (\alpha_k \rho_k q_k) -P_{\infty,k}\Big{)}.
$$

\begin{table}\centering
\caption{Parameters of the stiffened gas EOS for the CO\textsubscript{2} phases, used in the numerical test.}\vspace*{2ex}
\label{tab:1}
$\begin{array}{llll}  \hline
        &                     & \text{Liquid}& \text{Vapour} \\   \hline
\gamma  & [-]                 &  1.23        & 1.06          \\
P_\infty& [\mathrm{Pa}]       &  1.32~10^{8} &  8.86~10^{5}  \\
 q      & [\mathrm{J/kg}]     & -6.23~10^{5} & -3.01~10^{5}  \\
 c_v    & [\mathrm{J/kg \,K}] &  2440        & 2410          \\
\hline
\end{array}$ 
\end{table}

\begin{figure*}
\centering
 \includegraphics[width=0.95\textwidth]{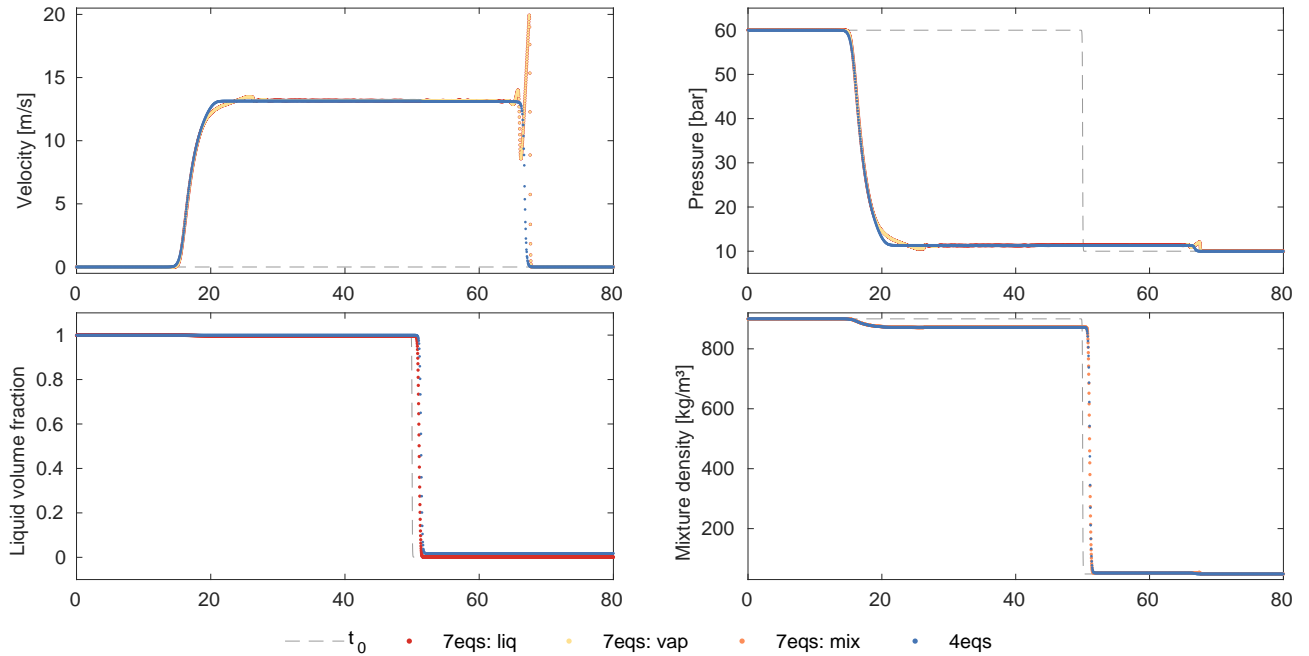}
 \caption{CO\textsubscript{2} de-pressurization test: numerical results obtained with the seven-equation model of Sec.~\ref{s:seveneqs}, referred to as \textsf{7eqs}, and the four-equation model of Sec.~\ref{s:foureqs}, referred to as \textsf{4eqs}, at the final time $t_\mathrm{F}=0.08~\mathrm{s}$.
In the first row, velocity and pressure are displayed. For the seven-equation model, cell values for both liquid and vapour are shown. 
In the second row, the volume fraction of the liquid and the mixture density, defined as $\rho = \alpha_1 \rho_1 + \alpha_2 \rho_2$, are shown. In all pictures, the initial solution is displayed as a dashed line.
In the simulation with the \textsf{7eqs} model, since the numerical discretization is only first-order accurate, the domain is split into 4000 cells; while in the simulation with the \textsf{4eqs} model, which is second-order accurate, 2000 cells are used.}
 \label{fig:01}
\end{figure*}

\section{Numerical results}\label{s:results}
In this section, we show the results obtained by both the pressure-based non-equilibrium BN-type model, presented in Sec.~\ref{s:seveneqs}, and the internal energy-based four-equation model, described in Sec.~\ref{s:foureqs}, on the two-phase
CO\textsubscript{2} de-pressurization test proposed in~\cite{LundAursand2012}.

The experiment considers pure CO\textsubscript{2}, initially at rest in a pipe of length $L=80~\mathrm{m}$.
The pipeline is split in two chambers, the left one ($x<50~\mathrm{m}$) contains liquid CO\textsubscript{2} at $P_\mathrm{L}=60~\mathrm{bar}$; the right one ($x>50~\mathrm{m}$) is filled with vapour at $P_\mathrm{R}=10~\mathrm{bar}$. The temperature is $T=273~\mathrm{K}$ everywhere.
The parameters of the stiffened gas EOS for liquid and vapour CO\textsubscript{2} are given in~\ref{tab:1}.

The diffuse interface methods assume that a small amount of all phases is present in every computational cells, so we impose a liquid volume fraction $\alpha_{1,\mathrm{L}}=0.999$ in the left part, and $\alpha_{1,\mathrm{R}}=0.001$ in the right part.
The results of the simulation at $t_\mathrm{F}=0.08~\mathrm{s}$ are shown in~\ref{fig:01}.

As the time evolves, the initial discontinuity generates a rarefaction wave propagating towards the left, within the liquid, and a shock propagating towards the right, across the vapour. The interface separating the fluids propagates as a contact discontinuity across which a jump in the (numerical) mixture density (defined as $\rho = \alpha_1 \rho_1 + \alpha_2 \rho_2$) takes place, but the pressure and velocity are continuous. Indeed, no spurious oscillations occur in these variables (first row of~\ref{fig:01}) at $x\approx 51~\mathrm{m}$, which is the location of the interfaces at the final time.
Unfortunately, the seven-equation model does generate velocity oscillations across the shock wave. As a possible explanation, we remind that this numerical method has been developed especially for low-Mach phase flows, i.e., in a regime where strong shocks are not expected.
However, we will further investigate this issue.

We observe a good agreement between our numerical results and the ones in~\cite{LundAursand2012}. The relaxation processes in the seven-equation model described in \ref{s:seveneqs} drive correctly the velocity and the pressure towards the equilibrium: as the pictures in the first row of~\ref{fig:01} show, the cell values of these variables are the same for liquid and vapour CO\textsubscript{2}.
These values, as well as the volume fraction and the mixture density, agree very well with the results achieved by the four-equation model, despite the profound differences in the modelling assumptions.

\section{Conclusions}\label{s:conc}
In this work, we have presented two different methods for the simulation of unsteady compressible two-phase or two-component flows.
In particular, we have focused on diffuse interface methods, which permit to describe each fluid according to its own equation of state.
We have shown how spurious oscillations across the interfaces that separate different fluids can be avoided by adopting primitive formulations, i.e., by solving the governing equations for the pressure or the internal energy, instead of the corresponding conservative variable, i.e. the total energy.

In the presented de-pressurization test of pure two-phase CO\textsubscript{2}, the two simulation strategies described here give similar results, although they rely on different modelling hypotheses, first of all, the mechanical and thermal equilibrium or dis-equilibrium between phases.
In particular, the physical constraint of continuous pressure and velocity across interfaces is correctly recovered in both simulations.

With this work, we wanted to stress the simple---and probably not new---observation that the simulation of compressible multiphase or multicomponent flows can be addressed in several ways but some similar conclusions can often be drawn, and used to implement new numerical methods.
Hence, a simulation strategy developed for a reduced model can serve as an important building block for more complex models, which sometimes are required to simulate unsteady multiphase flows, for which no equilibrium assumption can be made a priori.

\section*{Acknowledgement}
This publication has been produced with support from the \textit{NCCS Centre}, performed under the Norwegian research program Centres for Environment-friendly Energy Research (FME). B. Re and R. Abgrall acknowledge the following partners for their contributions: Aker Solutions, Ansaldo Energia, CoorsTek Membrane Sciences, Emgs, Equinor, Gassco, Krohne, Larvik Shipping, Norcem, Norwegian Oil and Gas, Quad Geometrics, Shell, Total, V\aa r Energi, and the Research Council of Norway (257579/E20).
We would also like to acknowledge Julien Carlier and Pietro Congedo for their support in developing the four-equation model, and Svend Tollak Munkejord (SINTEF, Norway) for the helpful discussions about multiphase CO\textsubscript{2} flows.

\bibliography{references} 

\begin{thebibliography}{10}

\bibitem{Saurel1999}
R.~Saurel and R.~Abgrall, ``{A Multiphase Godunov Method for Compressible
  Multifluid and Multiphase Flows},'' {\em J. Comput. Phys.}, vol.~150, no.~2,
  pp.~425--467, 1999.

\bibitem{Saurel2018}
R.~Saurel and C.~Pantano, ``{Diffuse-Interface Capturing Methods for
  Compressible Two-Phase Flows},'' {\em Annu. Rev. Fluid Mech.}, vol.~50,
  no.~1, pp.~105--130, 2018.

\bibitem{abgrall2018high}
R.~Abgrall, P.~Bacigaluppi, and S.~Tokareva, ``A high-order nonconservative
  approach for hyperbolic equations in fluid dynamics,'' {\em Computers \&
  Fluids}, vol.~169, pp.~10--22, 2018.

\bibitem{bacigaluppi2019high}
P.~Bacigaluppi, {\em High order fully explicit residual distribution
  approximation for conservative and non-conservative systems in fluid
  dynamics}.
\newblock PhD thesis, University of Zurich, 2019.

\bibitem{Baer1986}
M.~Baer and J.~Nunziato, ``{A two-phase mixture theory for the
  deflagration-to-detonation transition (DDT) in reactive granular
  materials},'' {\em Int. J. Multiph. Flow}, vol.~12, pp.~861--889, nov 1986.

\bibitem{Tokareva2010}
S.~Tokareva and E.~Toro, ``{HLLC-type Riemann solver for the Baer--Nunziato
  equations of compressible two-phase flow},'' {\em J. Comput. Phys.},
  vol.~229, no.~10, pp.~3573--3604, 2010.

\bibitem{Lund2012}
H.~Lund, ``{A Hierarchy of Relaxation Models for Two-Phase Flow},'' {\em SIAM
  J. Appl. Math.}, vol.~72, pp.~1713--1741, jan 2012.

\bibitem{Saurel2009}
R.~Saurel, F.~Petitpas, and R.~A. Berry, ``{Simple and efficient relaxation
  methods for interfaces separating compressible fluids, cavitating flows and
  shocks in multiphase mixtures},'' {\em J. Comput. Phys.}, vol.~228, no.~5,
  pp.~1678--1712, 2009.

\bibitem{Kapila2001}
A.~K. Kapila, R.~Menikoff, J.~B. Bdzil, S.~F. Son, and D.~S. Stewart,
  ``{Two-phase modeling of deflagration-to-detonation transition in granular
  materials: Reduced equations},'' {\em Phys. Fluids}, vol.~13, pp.~3002--3024,
  oct 2001.

\bibitem{downar1996non}
P.~Downar-Zapolski, Z.~Bilicki, L.~Bolle, and J.~Franco, ``{The Non-equilibrium
  Relaxation Model for One-dimensional Flashing Liquid Flow},'' {\em
  International journal of multiphase flow}, vol.~22, no.~3, pp.~473--483,
  1996.

\bibitem{LundAursand2012}
H.~Lund and P.~Aursand, ``{Two-Phase Flow of CO2 with Phase Transfer},'' {\em
  Energy Procedia}, vol.~23, pp.~246--255, 2012.

\bibitem{Abgrall2003}
R.~Abgrall and R.~Saurel, ``{Discrete equations for physical and numerical
  compressible multiphase mixtures},'' {\em J. Comput. Phys.}, vol.~186, no.~2,
  pp.~361--396, 2003.

\bibitem{Abgrall1996}
R.~Abgrall, ``{How to Prevent Pressure Oscillations in Multicomponent Flow
  Calculations: A Quasi Conservative Approach},'' {\em J. Comput. Phys.},
  vol.~125, no.~1, pp.~150--160, 1996.

\bibitem{Karni1996}
S.~Karni, ``{Hybrid Multifluid Algorithms},'' {\em SIAM J. Sci. Comput.},
  vol.~17, no.~5, pp.~1019--1039, 1996.

\bibitem{Re2019}
B.~Re and R.~Abgrall, ``{Non-equilibrium Model for Weakly Compressible
  Multi-component Flows: the Hyperbolic Operator},'' in {\em Non-Ideal
  Compressible-Fluid Dyn. Propuls. Power} (F.~{Di Mare}, A.~Spinelli, and
  M.~Pini, eds.), Springer, 2019.

\bibitem{Wenneker2002}
I.~Wenneker, A.~Segal, and P.~Wesseling, ``{A Mach-uniform unstructured
  staggered grid method},'' {\em Int. J. Numer. Methods Fluids}, vol.~40,
  pp.~1209--1235, nov 2002.

\bibitem{Re2020}
B.~Re and R.~Abgrall, ``{Pressure-based non-equilibirum model for the
  simulation of weakly compressible multiphase flows},'' {\em In preparation},
  2020.

\bibitem{Munz2003}
C.-D. Munz, S.~Roller, R.~Klein, and K.~J. Geratz, ``{The extension of
  incompressible flow solvers to the weakly compressible regime},'' {\em
  Comput. Fluids}, vol.~32, no.~2, pp.~173--196, 2003.

\bibitem{petsc-web-page}
``{PETSc Web page}.'' \url{http://www.mcs.anl.gov/petsc}, 2018.

\bibitem{Zein2010}
A.~Zein, M.~Hantke, and G.~Warnecke, ``{Modeling phase transition for
  compressible two-phase flows applied to metastable liquids},'' {\em J.
  Comput. Phys.}, vol.~229, no.~8, pp.~2964--2998, 2010.

\bibitem{bacigaluppi2020}
P.~Bacigaluppi, J.~Carlier, R.~Abgrall, M.~Pelanti, and P.~Congedo,
  ``Assessment of a non-conservative residual distribution scheme for solving a
  four-equation two-phase system with phase transition.''
\newblock in preparation.

\bibitem{abgrall2017design}
R.~Abgrall and P.~Bacigaluppi, ``Design of a second-order fully explicit
  residual distribution scheme for compressible multiphase flows,'' in {\em
  International Conference on Finite Volumes for Complex Applications},
  pp.~257--264, Springer, 2017.

\bibitem{BacigaluppiMood2018}
P.~Bacigaluppi, R.~Abgrall, and S.~Tokareva, ``{``A Posteriori Limited'' High
  Order and Robust Residual Distribution Schemes for Transient Simulations of
  Fluid Flows in Gasdynamics}.''
\newblock in (minor) revision.

\bibitem{Ricchiuto2010}
M.~Ricchiuto and R.~Abgrall, ``{Explicit {Runge-Kutta} Residual Distribution
  Schemes for Time Dependent Problems: Second Order Case},'' {\em J. Comput.
  Phys.}, vol.~229, no.~16, pp.~5653--5691, 2010.

\bibitem{Pelanti2014}
M.~Pelanti and K.~Shyue, ``{A Mixture-energy-consistent Six-equation Two-phase
  Numerical Model for Fluids with Interfaces, Cavitation and Evaporation
  Waves},'' {\em Journal of Computational Physics}, vol.~259, pp.~331--357,
  2014.

\bibitem{LeMetayer2004}
O.~{Le M{\'{e}}tayer}, J.~Massoni, and R.~Saurel, ``{{\'{E}}laboration Des Lois
  D'{\'{E}}tat D'Un Liquide Et De Sa Vapeur Pour Les Mod{\`{e}}les
  D'{\'{E}}coulements Diphasiques},'' {\em Int. J. Therm. Sci.}, vol.~43,
  no.~3, pp.~265--276, 2004.

\end{thebibliography}
\bibliographystyle{ieeetr}

\end{document}